\documentstyle[12pt, bezier]{article}
\def\beq{\begin{equation}}
\def\eeq{\end{equation}}
\def\hb{\hat \beta}
\def\ap#1#2#3 {Ann. Phys. (NY) {\bf#1} (19#2) #3}
\def\apj#1#2#3 {Astrophys. J. {\bf#1} (19#2) #3}
\def\apjl#1#2#3 {Astrophys. J. Lett. {\bf#1} (19#2) #3}
\def\app#1#2#3 {Acta. Phys. Pol. {\bf#1} (19#2) #3}
\def\ar#1#2#3 {Ann. Rev. Nucl. Part. Sci. {\bf#1} (19#2) #3}
\def\cpc#1#2#3 {Computer Phys. Comm. {\bf#1} (19#2) #3}
\def\err#1#2#3 {{\it Erratum} {\bf#1} (19#2) #3}
\def\ib#1#2#3 {{\it ibid.} {\bf#1} (19#2) #3}
\def\jmp#1#2#3 {J. Math. Phys. {\bf#1} (19#2) #3}
\def\ijmp#1#2#3 {Int. J. Mod. Phys. {\bf#1} (19#2) #3}
\def\jetp#1#2#3 {JETP Lett. {\bf#1} (19#2) #3}
\def\jpg#1#2#3 {J. Phys. G. {\bf#1} (19#2) #3}
\def\mpl#1#2#3 {Mod. Phys. Lett. {\bf#1} (19#2) #3}
\def\nat#1#2#3 {Nature (London) {\bf#1} (19#2) #3}
\def\nc#1#2#3 {Nuovo Cim. {\bf#1} (19#2) #3}
\def\nim#1#2#3 {Nucl. Instr. Meth. {\bf#1} (19#2) #3}
\def\np#1#2#3 {Nucl. Phys. {\bf#1} (19#2) #3}
\def\pcps#1#2#3 {Proc. Cam. Phil. Soc. {\bf#1} (#2) #3}
\def\pl#1#2#3 {Phys. Lett. {\bf#1} (19#2) #3}
\def\prep#1#2#3 {Phys. Rep. {\bf#1} (19#2) #3}
\def\prev#1#2#3 {Phys. Rev. {\bf#1} (19#2) #3}
\def\prl#1#2#3 {Phys. Rev. Lett. {\bf#1} (19#2) #3}
\def\prs#1#2#3 {Proc. Roy. Soc. {\bf#1} (19#2) #3}
\def\ptp#1#2#3 {Prog. Th. Phys. {\bf#1} (19#2) #3}
\def\ps#1#2#3 {Physica Scripta {\bf#1} (19#2) #3}
\def\rmp#1#2#3 {Rev. Mod. Phys. {\bf#1} (19#2) #3}
\def\rpp#1#2#3 {Rep. Prog. Phys. {\bf#1} (19#2) #3}
\def\sjnp#1#2#3 {Sov. J. Nucl. Phys. {\bf#1} (19#2) #3}
\def\spj#1#2#3 {Sov. Phys. JEPT {\bf#1} (19#2) #3}
\def\spu#1#2#3 {Sov. Phys. Usp. {\bf#1} (19#2) #3}
\def\zp#1#2#3 {Zeit. Phys. {\bf#1} (19#2) #3} 

\begin{document}
\begin{titlepage}
\begin{center}
{\Large \bf Theoretical Physics Institute \\
University of Minnesota \\}  \end{center}
\vspace{0.3in}
\begin{flushright}
TPI-MINN-96/14-T \\
UMN-TH-1508-96 \\
August 1996
\end{flushright}
\vspace{0.4in}
\begin{center}
{\Large \bf  Domain Walls are Diamagnetic\\}
\vspace{0.2in} 
{\bf M.B. Voloshin  \\ } 
Theoretical Physics Institute, University of Minnesota, Minneapolis, MN 
55455 \\ and \\ 
Institute of Theoretical and Experimental Physics, Moscow, 117259 \\[0.2in]

{\bf   Abstract  \\ }
\end{center}

It is shown that contrary to a recent claim in the 
literature$^{\cite{iwazaki}}$, the domain walls made of scalar field are 
diamagnetic due to presence of massless fermionic modes on the wall. The 
diamagnetism vanishes at high temperature.  Thus the domain walls could 
produce no effect on a primordial magnetic field in the early universe.

\end{titlepage} 

In a recent paper Iwazaki$^{\cite{iwazaki}}$ has considered the magnetic
properties of the gas of massless fermions bounded to a domain wall
in a theory with spontaneously broken discrete symmetry. The result is
claimed to be that the free energy of the gas in magnetic field with
the strength $B$ contains a negative term proportional to $-B^{3/2}$,
thus that the gas should exhibit spontaneous magnetization. If this were
true, the domain walls in the early universe could be a source of the
primordial magnetic field.

However, the calculations in Ref. 1 are not quite correct, and the
purpose of this note is to present a correct calculation of the magnetic
properties of the gas of massless fermions on the wall at zero
temperature as well as at finite temperature. The result of the present
calculation is that in fact the term in the free energy proportional to
$B^{3/2}$ is {\it positive} and goes to zero at high temperature $T^2
\gg e\, B$. Thus the domain walls are diamagnetic, and they can not
generate a primordial magnetic field in the early universe.

It can be reminded that a domain wall, located for definiteness
perpendicular to the  $z$ axis, corresponds to transition of a scalar
field $\phi (z)$ between two minima $v_1$ and $v_2$ of the scalar field
potential: $\phi (z=-\infty) = v_1$, $\phi (z=+\infty) = v_2$.
Furthermore, if there is a fermion field $\psi$ with a Yukawa coupling to
$\phi$ such that the fermion mass term $m(\phi)$ changes sign between 
$v_1$ and $v_2$ such fermion has a zero mode with
respect to motion in the $z$ direction$^{\cite{jr}}$. The fermion wave
function corresponding to this zero mode thus splits into the product
$\psi(t,x,y,z)=\chi(t,x,y) \, u(z)$ with the scalar function $u(z)$
given by $u(z) = {\rm const} \cdot \exp \left ( \pm \int^z m \, dz
\right )$ and the spinor $\chi$ being an eigenvector of the Dirac matrix
$i \, \gamma_3$: $i \, \gamma _3 \chi = \pm \chi$, where the upper
(lower) sign corresponds to $m(v_2)$ negative (positive). The spinor
condition on $\chi$ leaves only two components independent, thus the
motion of the zero-mode fermions along the wall is described by the
(2+1) dimensional Dirac equation for a two-component massless spinor.
Clearly, the solutions to that equation (on-shell particles)
have one degree of freedom for fermions and one for antifermions. In the
absence of an external magnetic field the spectrum of the free Dirac
operator is obviously labeled by the two components of the
two-dimensional momentum {\bf k}: $E(k)=\pm k$ (with $k=|${\bf k}$|$).
In the presence of a uniform magnetic field $B$ orthogonal to the
wall, the spectrum of the Dirac operator with the electric charge $e$ 
is found$^{\cite{iwazaki}}$ following the textbook method of
Landau$^{\cite{ll}}$ and is parametrized by one integer quantum number 
$n \ge 0$ with the degeneracy $e \, B/(2 \pi)$ per unit area of the
wall and the energy given by $E_n=\pm \sqrt{2 \, e \, B \, n}$.

The existence of the non-zero modes of the motion in the $z$ direction,
either non-localized or localized$^{\cite{mv}}$ at the wall, can be
ignored in calculating the magnetic and thermodynamic properties of
the gas of the zero-mode fermions as long as the characteristic energy 
scale in the problem, i.e. $\sqrt{e \, B}$ and/or the temperature $T$, is
much smaller than the mass gap given by the mass of the fermions in a
spatially uniform vacuum state $v_1$ or $v_2$. Throughout this letter
this is assumed to be the case and the contribution of non-zero modes 
to the partition function is completely ignored. 

The standard construction of the physical states of the fermion field
starts with defining the vacuum state, where the negative energy states
are occupied and the positive energy ones are vacant. In the considered 
here system of the
two-dimensional massless fermions this leaves a $(Z_2)^{N}$
degeneracy of the vacuum states corresponding to the two possibilities
for the occupation number at $n=0$ and the total number of such states
is $N={e \, B \over 2 \pi} \times Area$. The non-vacuum states, as
usual,
correspond to fermions: filled states with positive energy, and to
antifermions: holes in the states with negative energy. Accordingly, 
the free energy per unit area of the fermionic gas at a temperature 
$T=1/\beta$ in the  magnetic field $B$  can be written in the following 
form
\beq
F = F_- + F_+ + F_0 + E_{\rm vac}~~,
\label{fe}
\eeq
where 
\beq
F_\pm = - \beta^{-1} \, {e \, B \over 2 \pi}\, 
\sum_{n=1}^\infty \, \ln \, \left ( 1+ e^{-\beta
\sqrt{2 \, e \, B \, n}} \right )
\label{fpm}
\eeq 
is the free energy associated with the real gas of fermions and
antifermions\footnote{The fermion number can be freely exchanged with
the non-zero modes, hence there is no chemical potential in
eq.(\ref{fpm})$^{\cite{iwazaki}}$.}. The term $F_0$ in eq.(\ref{fe}) is 
associated with the degeneracy of the vacuum state:
\beq
F_0=-\beta^{-1} \, {e \, B \over 2 \pi}\, \ln \,2 ~~.
\label{f0}
\eeq
Finally, $E_{\rm vac}$ is the energy per unit area of the wall of any of 
the degenerate vacuum states\footnote{It is this vacuum energy term
which is missing in the corresponding calculation of Ref. 1. Also 
the Euler-Maclaurin expansion for the sum in
eq.(\ref{fpm}) is unjustifiably truncated, which has lead to a wrong
result there. In fact this expansion is of little help in calculating 
the sum because of the root singularity of the summand at $n=0$.}.
Clearly, $E_{\rm vac}$ gives the free energy at $T=0$.

Ignoring the divergence of the sum, one might write the vacuum energy 
as the sum over the energies of the occupied negative-energy eigenstates:
\beq
E_{\rm vac}=-{e \, B \over 2 \pi} \, \sum_{n=1}^\infty \, 
\sqrt{2 \, e \, B \, n} = -{e \, B \over 2 \pi} \, \sqrt{2 \,e \, B } \, \zeta
\left ( - {1 \over 2} \right ) \approx 0.04679 \, 
\left ( e \, B \right )^{3/2}~~,
\label{e0}
\eeq
where the finite answer for the divergent sum is written in terms of the 
standard $\zeta(s)$ function analytically continued below its pole at
$s=1$. 
It turns out that this formal manipulation gives the correct expression
for the dependence of the vacuum energy on the field strength $B$, while
the overall vacuum energy is of course infinite. In view of this
divergence a somewhat more elaborate consideration of the vacuum energy
is due, and this consideration, presented in the following lines also
introduces an appropriate method for calculating the sums involved in
this problem\footnote{I am thankful to Igor Aleiner for reminding me
this summation technique, which  in fact was used by Landau in similar 
problems and which originally goes back to Poisson.}. 

First, in order to make tractable the sum
over the infinite number of negative energy states, it has to
be regularized. A gauge invariant regulator factor should depend on a
gauge invariant quantity, which in this case is naturally the level
energy itself. We chose here the regulator factor in the exponential
form: $\exp (- \epsilon \, E_n^2)$ with $\epsilon$ being the regulator
parameter. Thus we write the regularized vacuum energy in the form
\beq
E^{(r)}_{\rm vac}= -{e \, B \over 2 \pi} \, \sum_{n=1}^\infty \, 
\sqrt{2 \, e \, B \, n} \, \exp \left ( - \epsilon \, 2 \, e \, B \, 
n \right )~~.
\label{er}
\eeq
The method of summation to be used here is based on the identity valid
for any smooth function $f(x)$:
\beq
\sum_{n=1}^\infty \, f(n) = \int_\delta^\infty \, f(x) \, 
\sum_{n=-\infty}^\infty \, \delta(x-n) \, dx = \sum_{m=-\infty}^\infty \,
\int_\delta^\infty \, f(x) \, \exp( 2 \, \pi \, i \, m \, x) \, dx~~,
\label{id}
\eeq 
where $\delta$ is any number such that $0 < \delta <1$. Notice that the
sum in the second expression goes over $n$ from $-\infty$ to $+\infty$.
The identity is still valid since the terms with $n \le 0$ are
identically zero. Applying this identity to the sum in eq.(\ref{er}) and
noting that in this case in fact $\delta$ can be set equal to zero,
since the term in the sum with $n=0$ is vanishing, one can write
\beq
E^{(r)}_{\rm vac}= -{e \, B \over 2 \pi} \, \sum_{m=-\infty}^\infty \,
\int_0^\infty \, \sqrt{2 \, e \, B \, x} \, 
\exp \left ( - \epsilon \, 2 \, e \, B \, x
 \right )
\, \exp( 2 \, \pi \, i \, m \, x) \, dx~~.
\label{erx}
\eeq
For each $m$ the integral in this equation is clearly that for the
$\Gamma$ function: $\Gamma(3/2)$. The term with $m=0$ diverges in the
limit $\epsilon \to 0$, while all the rest terms of the sum are finite.
Thus writing separately the singular term and grouping together the
terms with symmetric values of $m$ one finds for $\epsilon \to 0$
\beq
E^{(r)}_{\rm vac} = -{\sqrt{\pi} \over 8 \, \pi \, \epsilon^{3/2}} +
{e \, B \over 2 \pi} \, \sqrt{2 \, e \, B} \, {\zeta(3/2) \over 4 \,
\pi}~~.
\label{evf}
\eeq
The singular term in this expression does not depend on $B$ and is
exactly equal to the energy of the free vacuum regularized in the same
way:
\beq
E^{(r)}_{\rm vac} (0) = -\int \, k \, e^{-\epsilon \, k^2} \, { d^2k
\over (2 \, \pi)^2} =  -{\sqrt{\pi} \over 8 \, \pi \, \epsilon^{3/2}}~~,
\label{ev00}
\eeq
while the second term in eq.(\ref{evf}), which is finite and describes
the dependence of $E_{\rm vac}$ on the magnetic field is exactly equal
to that in eq.(\ref{e0}) due to the identity $\zeta(-1/2) = -\zeta(3/2)
/(4 \pi)$, which is a consequence of a general relation$^{\cite{gr}}$
between $\zeta(s)$ and $\zeta(1-s)$.

The term with $B^{3/2}$ describing the dependence of the vacuum energy
on $B$ in eq.(\ref{ev00}) or in eq.(\ref{e0}) is positive. 
This corresponds to
diamagnetism of the wall at zero temperature with a singular at $B \to 0$
diamagnetic susceptibility. The temperature dependence
is given by the terms $F_\pm$ and $F_0$ in the free energy in
eq.(\ref{fe}). Following Ref.1, we consider here the free energy at a
large temperature, i.e. in the limit $\hb \equiv \beta \, \sqrt{2 \, e
\, B} \ll 1$ (but still the temperature is much less than the mass gap
in order to ensure the irrelevance of the non-zero modes). According to
eqs.(\ref{fe} - \ref{f0}), the
temperature dependent part of the free energy is given by
\beq
F_+ +F_- + F_0 = -\beta^{-1} \, { e \, B \over \pi} \left [
\sum_{n=1}^\infty \, \ln \left (1+ e^{-\hb \, \sqrt{n}} \right ) + {1
\over 2} \, \ln 2 \right ]~~.
\label{ft}
\eeq
In order to find the expansion of the sum in this expression in powers
of $\hb$ at small $\hb$ we apply to the sum the identity (\ref{id}).
Writing separately the Fourier harmonic with $m=0$ and grouping together
the harmonics with symmetric $m$, one finds
\begin{eqnarray}
&&\sum_{n=1}^\infty \, \ln \left (1+ e^{-\hb \, \sqrt{n}} \right ) =
\nonumber \\
&&\int_\delta^\infty \, \ln \left (1+ e^{-\hb \, \sqrt{x}} \right ) \, dx
+ \sum_{m=1}^\infty \, \int_\delta^\infty \, \ln \left (1+ e^{-\hb \,
\sqrt{x}} \right ) \, \left ( e^{2 \, \pi \, i \, m \, x} + e^{-2 \, \pi
\, i \, m \, x} \right ) \, dx~~.
\label{sf}
\end{eqnarray}
The first integral in the limit $\delta \to +0$ gives the singular in
$\hb$ term ${3 \over 2} \, {\zeta(3) \over \hb^2}$, which upon
substitution in eq.(\ref{ft}) reproduces the field independent free
energy of a free fermion gas. The rest of the terms of the expansion in
powers of $\hb$ are obtained by the Taylor expansion of the integrand in
the second term in eq.(\ref{sf}) with subsequent separate integration 
and summation over $m$ in each term of the expansion in $\hb$. The
parameter $\delta$ can be set to zero before integration in all the
terms  of the Taylor expansion except the first one, where keeping
$\delta$ small but finite ensures the convergence
of the sum over $m$. The calculation of this first, constant, term is
thus reduced to the formula
\beq
\lim_{\delta \to +0} \, \sum_{m=1}^\infty \, \int_\delta^\infty 
\, \left ( e^{2 \, \pi \, i \, m \, x} + e^{-2 \, \pi
\, i \, m \, x} \right ) \, dx = -{1 \over 2}
\label{i0}
\eeq
and the higher terms of the Taylor expansion in $\hb$ 
are found from the generic formula
\beq
\sum_{m=1}^\infty \, \int_0^\infty \, x^{p/2}
\, \left ( e^{2 \, \pi \, i \, m \, x} + e^{-2 \, \pi
\, i \, m \, x} \right ) \, dx = - { \sin(\pi \, p /4) \over \pi} \,
\left ( 2 \, \pi \right ) ^{-p/2} \, \Gamma \left ( {p \over 2} + 1
\right ) \, \zeta \left ( {p \over 2} +1 \right )
\label{ig}
\eeq
for $p > 0$. (Notice that eq.(\ref{i0}) can also be found by taking in 
this formula the limit $p \to 0$.)
In deriving these formulas the oscillating behavior of the integrand is
dealt with by temporarily introducing the damping factor $\exp(-\epsilon
\, x)$ and taking $\epsilon \to +0$ in the end: this is a legitimate
procedure, since it does not alter the original integral in
eq.(\ref{sf}). 

It is satisfying to see, using eq.(\ref{i0}), that upon substitution in 
eq.(\ref{ft}) the constant, $\hb$ independent, term in the sum of 
eq.(\ref{ft}) exactly cancels the contribution $F_0$ to the free energy
of the vacuum degeneracy factor. Thus, as expected on general grounds,
no term linear in $B$ arises in the free energy at any
temperature. Therefore, using the formula in eq.(\ref{ig}) for $p=1$ 
one finds  in the high
temperature limit the expansion for the thermal part of the free energy
up to the first $B$ dependent term as
\beq
F_+ +F_- + F_0 = -{3 \over 4} \, {\zeta(3) \over \beta^2} -
{e \, B \over 2 \pi} \, \sqrt{2 \, e \, B} \, {\zeta(3/2) \over 4 \,
\pi} + O \left ( \beta \, (e \, B)^2 \right )~~.
\label{ht}
\eeq
One can readily see that the term with $B^{3/2}$ in this expression
exactly cancels in the sum in eq.(\ref{fe}) the corresponding
contribution of the vacuum energy, given by eq.(\ref{evf}). Thus one
concludes that the singular at $B \to 0$ diamagnetic behavior of the
fermion gas at the wall vanishes at high temperature, as could be
expected on general grounds.  

It is a simple exercise to calculate numerically the sum in
eq.(\ref{ft}) for arbitrary $\hb$. However in view of the absence of any
dramatic phenomena, like the previously claimed$^{\cite{iwazaki}}$
ferromagnetism of the fermion gas, this calculation is of a little
practical interest.

As discussed above, the calculation here is restricted to the situation
where the temperature is much smaller than the mass gap for the
fermions. If this restriction is relaxed, one has to solve the problem
for the full spectrum$^{\cite{mv}}$ of the fermion 
scattering states in the presence of the domain wall in (3+1)
dimensions. In this case however, the effect of the wall on the magnetic
properties of the field system is small, and also no dramatic effects
can be expected. 

Thus we conclude that the domain walls,  
had they existed in the early universe
in spite of their undesirable cosmological effects pointed out in the
literature$^{\cite{zko}}$, would not have produced any significant 
impact on a large-scale primordial magnetic field.

I am thankful to I. Aleiner, L. Glazman, A. Larkin and A. Vainshtein for
enlightening discussions. This work is supported, in part, by the DOE 
grant DE-AC02-83ER40105.

\end{document}